\documentclass[twocolumn,a4paper,superscriptaddress,floatfix,showpacs]{revtex4}
\usepackage{amsmath,amssymb,epsfig,color,textcase}

\begin{document}

\title{The nature of the ferromagnetic ground state in the Mn$_4$ molecular magnet.}

\author{S.V.~Streltsov}
\affiliation{Institute of Metal Physics, S.Kovalevskoy St. 18, 620990 Ekaterinburg, Russia}
\email{streltsov@imp.uran.ru}
\author{Z.V.~Pchelkina}
\affiliation{Institute of Metal Physics, S.Kovalevskoy St. 18, 620990 Ekaterinburg, Russia}
\affiliation{Ural Federal University, Mira St. 19, 620002 Ekaterinburg, Russia}
\author{D.I.~Khomskii}
\affiliation{II. Physikalisches Institut, Universit$\ddot a$t zu K$\ddot o$ln,
Z$\ddot u$lpicher Stra$\ss$e 77, D-50937 K$\ddot o$ln, Germany}
\author{N.A.~Skorikov}
\affiliation{Institute of Metal Physics, S.Kovalevskoy St. 18, 620990 Ekaterinburg, Russia}
\author{A.O.~Anokhin}
\affiliation{Institute of Metal Physics, S.Kovalevskoy St. 18, 620990 Ekaterinburg, Russia}
\author{Yu.N.~Shvachko}
\affiliation{Institute of Metal Physics, S.Kovalevskoy St. 18, 620990 Ekaterinburg, Russia}
\author{M.A.~Korotin}
\affiliation{Institute of Metal Physics, S.Kovalevskoy St. 18, 620990 Ekaterinburg, Russia}
\author{V.I.~Anisimov}
\affiliation{Institute of Metal Physics, S.Kovalevskoy St. 18, 620990 Ekaterinburg, Russia}
\affiliation{Ural Federal University, Mira St. 19, 620002 Ekaterinburg, Russia}
\author{V.V.~Ustinov}
\affiliation{Institute of Metal Physics, S.Kovalevskoy St. 18, 620990 Ekaterinburg, Russia}


\date{\today}

\begin{abstract}
Using {\it ab initio} band structure and model calculations we studied magnetic
properties of one of the Mn$_4$ molecular magnets (Mn$_4$(hmp)$_6$), where two types
of the Mn ions exist: Mn$^{3+}$ and Mn$^{2+}$. The direct calculation of the
exchange constants in the GGA+U approximation shows that in contrast to a
common belief the strongest exchange coupling is not between two Mn$^{3+}$ ions
($J_{bb}$), but along two out of four exchange paths connecting Mn$^{3+}$ 
and Mn$^{2+}$ ions ($J_{wb}$). Within the perturbation theory we performed 
the microscopic analysis of different contributions to the exchange constants, 
which allows us to establish the mechanism for the largest ferromagnetic 
exchange. 
In the presence of the charge order the lowest in energy virtual excitations, 
contributing to the superexchange, will be not those across the Hubbard gap $\sim U$, 
but will be those between the Mn$^{3+}$ and Mn$^{4+}$ ions, which cost much smaller 
energy $V$ ($\ll U$).
Together with strong Hund's rule coupling and specific orbital order this leads
to large ferromagnetic exchange interaction for two out of four 
Mn$^{2+}$--Mn$^{3+}$ pairs.
\end{abstract}

\maketitle

\section{Introduction \label{intro}}
The magnetic materials are so ubiquitous and essential for a plenty of devices 
used in the every day life that their properties are just taken for granted. 
The refrigerator magnets, medical implants, loudspeakers, magnetic resonance
imaging scanners, magneto-optical disks, electrical motors and generators, etc. - 
all these devices use permanent or nonpermanent magnets. Yet such an ancient and 
customary phenomenon as magnetism offers plenty of amazing new aspects. One of them is 
the molecule based magnetism discovered about two decades ago.~\cite{Sessoli1993}

It was revealed that single-molecule magnets (SMMs) and single-chain magnets having 
large spin and strong easy-axis anisotropy can self-assemble into 2D and 3D networks. 
This gives a great hope that one bit of information could be stored on a single molecule.~\cite{Jeon2012}
The new multidisciplinary field developed at the interface of chemistry and solid state 
physics was triggered by the finding that SMMs, exhibit both classical and 
exotic quantum magnetic properties.~\cite{Gatteschi2006} 
Challenged by the promising molecular spintronics and quantum computing applications
sophisticated SMMs based not only on 3$d$ transition metals but also on the 4$d$ and 
even on the lanthanide and actinide elements were 
developed.~\cite{Jiang2012,Wang2011,vardeny2010organic,winpenny2012molecular}
The theoretical studies have mostly concentrated on the description
of the resonant tunneling experiments,~\cite{Prokofev1998,Michalak2010} $ab$ $initio$ 
simulations,~\cite{RostamzadehRenani2012,Boukhvalov2007,Islam2010} and investigation of the role 
of the correlation effects.~\cite{Boukhvalov2008}

The single-molecule magnets consist of a core and bridging polynuclear complexes.
The physical insight into the magnetic interactions within the core is essential for both 
fundamental and technological development. At present the values of the exchange constants 
in SMMs are typically estimated by the fitting of the experimental magnetic susceptibility to 
a some solution of the Heisenberg model. There are number of deficiencies in such
a strategy mainly related with the large number of fitting parameters and
sometimes with an arbitrariness in the choice of the model Hamiltonian, which in 
general must include not only different direct and superexchange interactions,
but also magneto-crystalline anisotropy, Dzyaloshinskii-Moriya terms etc.

\begin{figure}[b!]
 \centering
 \includegraphics[clip=false,width=0.4\textwidth]{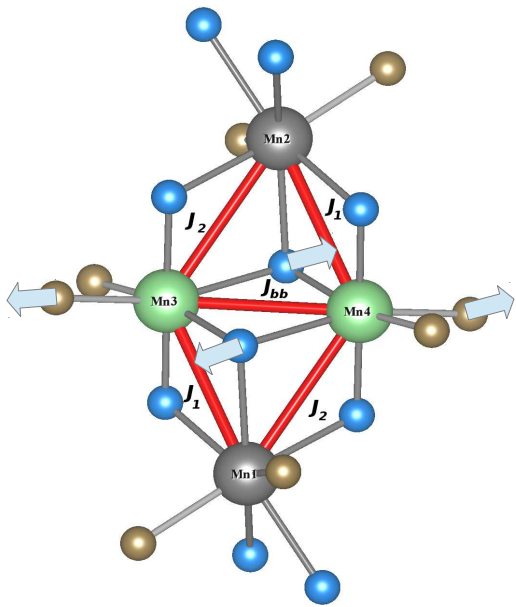}
\caption{\label{cryst.str}(color online). A fragment of the Mn$_4$ crystal structure
is shown. The Mn$^{2+}$ ions (Mn1 and Mn2) are shown as grey, Mn$^{3+}$ 
(Mn3 and Mn4) as green, O as blue, and N as brown
balls. Arrows show the direction of the Jahn-Teller elongation of the Mn$^{3+}$O$_6$
octahedra. The path for the ``body-body'' exchange coupling is labeled as 
$J_{bb}$, while two different the ``wing-body'' exchange integrals are
denoted via $J_1$ and $J_2$.
}
\end{figure}

In the present paper we calculate the exchange constants in one of the Mn$_4$ molecular
magnets ([Mn$_4$(hmp)$_6$(NO$_3$)$_2$FeNO(CN)$_5$]$\cdot$4CH$_3$CN)  
with two types of Mn ions (Mn$^{2+}$ and Mn$^{3+}$) having ``butterfly'' geometry. 
Experimentally it is established that the ground state of this molecule is ferromagnetic.
It is commonly assumed that there are two exchange interactions in the Mn$_4$ molecule 
magnets: $J_{bb}$ (body-body) between two Mn$^{3+}$ ions and $J_{wb}$ (wing-body) between 
Mn$^{2+}$ and Mn$^{3+}$ ions (see Fig.~\ref{cryst.str}, where $J_1$ and $J_2$ are two 
different $J_{wb}$), both of which are ferromagnetic with the first being much larger 
than the second.~\cite{Roubeau2008} The direct calculations presented in this paper 
reveal that this accepted picture is in fact incorrect: the dominant ferromagnetic 
exchange is not $J_{bb}$, but $J_{wb}$, with two inequivalent $J_{wb}$ being 
very different. The magnetic susceptibility obtained by the exact diagonalization 
method with the use of the calculated exchange integrals agrees with experimental data.

The detailed microscopic analysis shows that there are many exchange processes in the 
Mn$_4$(hmp)$_6$ molecular magnet, which partially compensate each other, but
its magnetic properties are mainly defined by two features of this system.
First of all, the Jahn-Teller distortions leads to a specific orbital
order, which in turn makes two exchange paths between Mn$^{2+}$ and Mn$^{3+}$ ions
inequivalent. Second, the charge order strongly modifies the exchange
processes between Mn ions of different valences and favors ferromagnetic exchange 
coupling. The results obtained allows not only to describe magnetic
properties of the Mn$_4$(hmp)$_6$ molecular magnet, but can be applied to other systems, 
including transition metal oxides, 
with a charge ordered ground state. In conclusion, we suggest some recipes to increase the 
value of the ferromagnetic exchange in the Mn$_4$ molecule magnets on the basis of the 
microscopic model developed in the present study.
We believe that the strategy presented here will be useful for other SMMs.

\section{Crystal structure and calculation details\label{details:sec}}

The crystal structure for system 
[Mn$_4$(hmp)$_6$(NO$_3$)$_2$(H$_2$O)$_2$](ClO$_4$)$_2$$\cdot$4H$_2$O,
abbreviated as Mn$_4$(hmp)$_6$ in what follows, was taken from Ref.~\onlinecite{Kushch2011}.
The main building block of Mn$_4$(hmp)$_6$ is the core consisting of four Mn ions
(two Mn$^{2+}$ and two Mn$^{3+}$)
and surrounding ligands shown in Fig.~\ref{cryst.str}. Since Mn$^{3+}$
are Jahn-Teller active ions, the ligand octahedra surrounding them
are strongly distorted. It was shown in Ref.~\onlinecite{Khomskii2003}
that in the case of identical ligands the elastic interaction favors
a parallel order of the elongated Mn-O bonds in the Jahn-Teller octahedra 
having a common edge. Surprisingly this is also the case in Mn$_4$(hmp)$_6$.

Note right away that for the $J_2$ bonds (Mn1-Mn4 and Mn2-Mn3) the
long Mn$^{3+}$-O bonds, shown in Fig.~\ref{cryst.str} by broad arrows, lie in the 
plane of corresponding Mn1-O-Mn4-O and Mn2-O-Mn3-O  plaquettes, whereas 
for the bonds $J_1$ (Mn1-Mn3 and Mn2-Mn4) these long bonds are perpendicular 
to such plaquettes. As we show below, this will finally give very different 
exchange constants $J_1$ and $J_2$. 

The band structure calculations were performed within the Density Functional
Theory (DFT). This type of the calculations was proven to provide
adequate description of many organic compounds including mixed-valence
systems,~\cite{Barone2002,Streltsov2012a,Boukhvalov2007} while some restrictions 
related with the computation of the low spin states has to be mentioned.~\cite{Bencini2009}
The projector augmented wave (PAW) method as implemented in
the Vienna ab initio simulation package (VASP) was used.~\cite{Kresse1996}
The exchange-correlation potential was chosen to be in
Perdew-Burke-Ernzerhof (PBE) form.~\cite{Perdew1996}
Non-spherical contributions from the gradient corrections inside the 
PAW spheres were included in the calculation scheme.
In order to take into account strong electronic correlations on the Mn sites 
the GGA+U approximation (the generalized gradient approximation taking
into account on-site $U$ Hubbard correction) was applied~\cite{Anisimov1997} with the on-site Coulomb 
repulsion parameter $U$=4.5~eV and the intra-atomic Hund's rule exchange 
$J_H$=0.9 eV.~\cite{Streltsov2008} The spin-orbit coupling was not taken into
account in the present calculations, so that the effects related
to this interaction (such as, e.g., the single-ion anisotropy) 
were not considered.

The mesh of 8 {\bf k}-points was used in the course 
of the self consistency. The integration of the bands was performed by the
tetrahedron method with the Bl\"ochl corrections.~\cite{Blochl1994a}

The magnetic susceptibility was calculated using the exact diagonalization
technique of the Heisenberg model 
\begin{eqnarray}
\label{Heisenberg}
\nonumber
\hat H &=& 2J_{bb} \hat {\vec S}_3 \hat {\vec S}_4 
+2J_1 (\hat {\vec S}_1 \hat {\vec S}_3 + \hat {\vec S}_2 \hat {\vec S}_4) \\
&+&2J_2 (\hat {\vec S}_1 \hat {\vec S}_4 + \hat {\vec S}_2 \hat {\vec S}_3)
\end{eqnarray}
implemented in the ALPS package.~\cite{ALPS} The exchange constants $J_{bb}$, $J_1$,
and $J_2$ were calculated from the total energies of four different
magnetic configurations as discussed in Sec.~\ref{results}. The numeration of the spins 
and Mn ions in Eq.~\eqref{Heisenberg} and Fig.~\ref{cryst.str} is the same. 
\begin{figure}[t!]
 \centering
 \includegraphics[clip=false,width=0.35\textwidth]{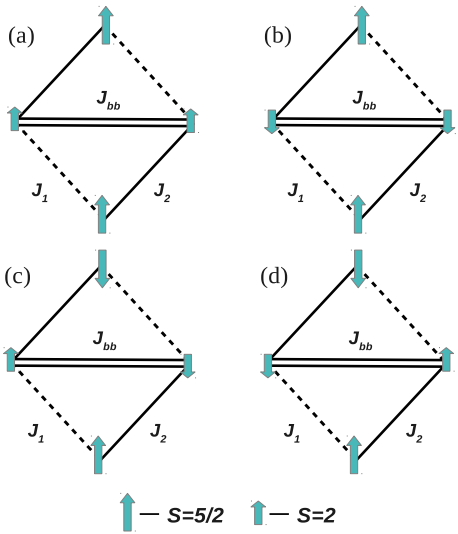}
\caption{\label{magn.str}(color online). Magnetic configurations
used for the exchange coupling calculations.
}
\end{figure}

%%%%%%%%%%%%%%%%%%%%%%%%%%%%%%%%%%%%%%%%%%%%%%%%%%%%%%%%%%%%%%%%%%%%%%%%%%%%%%%%%
\section{\label{results}Results of the calculations and comparison with
experiment}
\begin{figure}[t!]
 \centering
 \includegraphics[clip=false,width=0.5\textwidth]{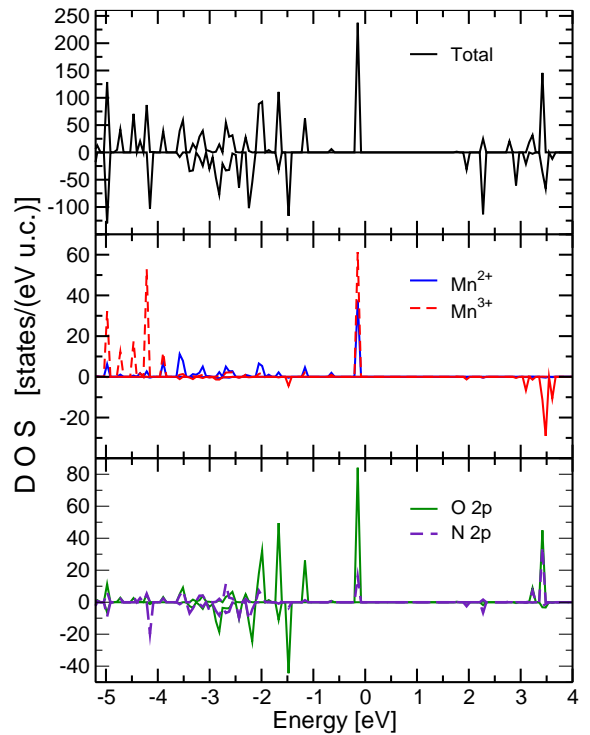}
\caption{\label{DOS}(color online). The total and partial density of states (DOS)
of the Mn$_4$ obtained in the ferromagnetic GGA+U calculation. The positive (negative)
values correspond to the spin up (down) states. The Fermi energy is in zero.}
\end{figure}

The total and partial density of states (DOS) obtained in the GGA+U
calculation for the fully ferromagnetic order
(Fig.~\ref{magn.str}a) are presented in Fig.~\ref{DOS}. One may see that
the top of the valence band is formed mostly by the O $2p$ states with admixture
of the N $2p$, Mn$^{3+}$ and Mn$^{2+}$ $3d$ states. The lower Hubbard
bands corresponding to the $3d$ states of the Mn$^{3+}$ and Mn$^{2+}$ ions
are in the range from -5 eV to $\sim$-2 eV. The magnetic moments
on the Mn$^{2+}$ and Mn$^{3+}$ were found to be 4.6~$\mu_B$
and 3.7~$\mu_B$, respectively. The deviations from the ionic
values (5~$\mu_B$ and 4~$\mu_B$) are related with the hybridization  and covalency effects, 
which result in a transfer of a part of the spin density to the ligands.
Similar effects were found in many other systems based on the transition
metal ions.~\cite{Streltsov2008a,Streltsov2013} The lowest total 
energy corresponds to the magnetic configuration, when all Mn ions
are ordered ferromagnetically.

As was already mentioned, typically only two exchange constants $J_{bb}$ and 
$J_{wb}$ are considered in the analysis of the magnetic properties of the Mn$_4$ molecular 
magnets.~\cite{Jerzykiewicz2010,Kushch2011}. In Mn$_4$(hmp)$_6$ $J_{wb}$ is the 
exchange constant between Mn$^{2+}$ and Mn$^{3+}$, and $J_{bb}$ between two 
Mn$^{3+}$. However, a close inspection of the available crystal 
structure~\cite{Kushch2011} shows that there are
two different distances between the Mn$^{2+}$ and Mn$^{3+}$ ions: $d_1(Mn^{2+}-Mn^{3+})$
=3.28~\AA, and $d_2(Mn^{2+}-Mn^{3+})$=3.34~\AA, which may result in two
different exchange couplings $J_{wb}$. In order to check this hypothesis
we calculated total energies of four magnetic structures presented in Fig.~\ref{magn.str},
which allows to extract three different exchange constants: 
$J_{bb}^{Calc}=-0.3$~K,  $J_{1}^{Calc}=-6.3$~K, and $J_{2}^{Calc}=-0.5$~K, all of them are 
ferromagnetic. Here $J_{1}$ and $J_{2}$ are two inequivalent $J_{wb}$
exchange couplings, which are shown in Fig.~\ref{cryst.str} and~\ref{magn.str}.
This result is quite in contrast to the common opinion that $J_{bb}$ must be
much larger than any of $J_{wb}$. 

The conventional assumption ($|J_{bb}| > |J_{wb}|$) was justified by 
two arguments. First of all the Mn-Mn distance along the
diagonal (3.20 \AA~in the present system) is much smaller 
than along the edges of the Mn$_4$ rhombus. Since the exchange constant
in the simplest case~\cite{Goodenough}
\begin{equation}
\label{t2U}
J \sim \frac {t_{dd}^2}U
\end{equation}
and the hopping integral for the $d$ orbitals ($t_{dd}$) is supposed to be inversely 
proportional to the distance between ions $t_{dd} \sim 1/r^5$,~\cite{Harrison1999}
this viewpoint seems to be justified. However the expression for the 
superexchange interaction in a particular situation can be quite different 
from Eq.~\eqref{t2U}. The detailed analysis performed in Sec.~\ref{Anal:sec} 
shows that for the given system $J_1$ is indeed expected to be
ferromagnetic and its absolute value is much larger than $|J_2|$ and possibly larger than
$|J_{bb}|$.

Second, the values of the exchange constants are typically extracted from the 
fitting of the experimental magnetic susceptibility $\chi(T)$ by the theoretical
curve obtained from the solution of the Heisenberg model.~\cite{Dey2012,Kushch2011} 
There are three ($g$, $J_{bb}$, and $J_{wb}$) or even four ($g$, $J_{bb}$, $J_1$, 
$J_2$) fitting parameters, which results in an arbitrariness of this
procedure. Instead of the fitting we first performed a direct calculation
of the temperature dependence of $\chi T$ (as described in Sec.~\ref{details:sec}) 
with the exchange constants obtained in the band structure calculations.
The single variable parameter ($g-$factor) was chosen to
fit the high-temperature tail of $\chi T$ and it was found to be equal 1.86. 

\begin{figure}[t!]
 \centering
 \includegraphics[clip=false,angle=270,width=0.5\textwidth]{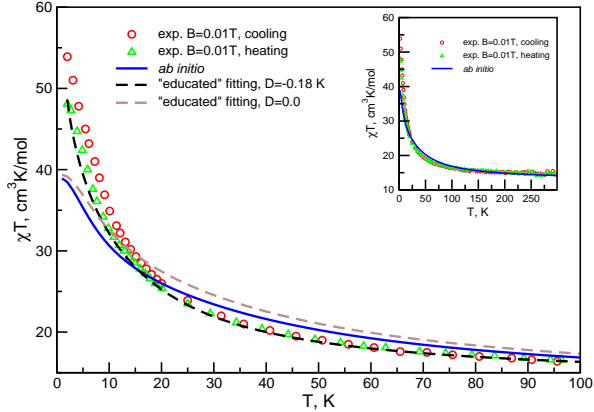}
\caption{\label{chiT}(color online). Temperature dependence of the
experimental and calculated $\chi T$. All theoretical curves were obtained by
exact diagonalization of the isotropic exchange interaction described by 
Hamiltonian~\eqref{Heisenberg} with the exchange constants, obtained in the GGA+U 
approximation, or using fitting taking into account or
neglecting by the single ion anisotropy ($D$). $\chi$ is the molar
susceptibility.}
\end{figure}

One may see from Fig.~\ref{chiT} that there is a reasonable agreement between calculated and
experimental curves for B=0.01 T. One may improve this agreement in the range from
20 to 100 K performing the fit to experimental data using calculated
in the GGA+U exchange constants as a starting point. This yields 
$J_{bb}^{fit}=-0.01$~K,  $J_{1}^{fit}=-6.5$~K, $J_{2}^{fit}=-0.2$~K, and $g=1.87$~K.
Further improvement can be achieved by adding the single ion anisotropy (via $DS_z^2$ 
terms for the Jahn-Teller Mn$^{3+}$ ions) into the fitting scheme.
This gives 
$J_{bb}^{fit,D}=-0.3$~K,  $J_{1}^{fit,D}=-3.6$~K, $J_{2}^{fit,D}=-0.4$~K, $D^{fit,D}=-0.18$~K, and $g=1.94$~K. 
Thus while $J$ obtained in this ``educated'' fitting differ from those 
calculated in the GGA+U, the main result is the same --  $J_{1}$ is the largest exchange constant.
The difference between theoretical and experimental 
data for $T<$10~K is related with experimental features of the measurements and
Mn$_4$(hmp)$_6$ samples, as discussed in Sec.~\ref{exp}.

\subsection{\label{exp} Low temperature behavior of $\chi(T) T$
- experimental difficulties}

\begin{figure}[t!]
 \centering
 \includegraphics[clip=false,width=0.4\textwidth]{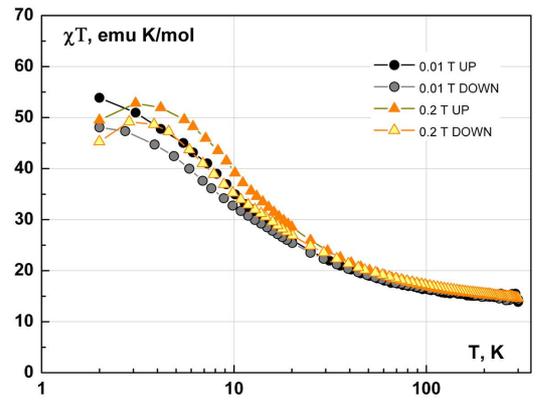}
\caption{\label{chiT-exp}(color online). 
The experimental $\chi T$ vs. T data obtained for polycrystalline sample in different magnetic fields 
B = 0.01 and  0.20 T for the cooling (DOWN) and heating (UP) regimes. The solid lines 
are guided by eye.
}
\end{figure}

Fig.~\ref{chiT-exp} discloses the details of measurements for the dried (aged) polycrystalline sample 
[Mn$_4$(hmp)$_6$(NO$_3$)$_2$FeNO(CN)$_5$]·4CH$_3$CN in Ref.~\onlinecite{Kushch2011}. 
Due to low spin state, $S=0$, 
the Fe$^{2+}$ ion does not contribute to the total paramagnetic response. Temperature 
dependencies for the product $\chi T$ are obtained at magnetic fields B=0.01 and 0.20~T. 
The measurements are performed at cooling from 300 K down to 2.0 K and at heating to ambient 
conditions. The temperature axis in Fig.~\ref{chiT-exp} is in logarithmic scale for better view 
of low temperature behavior. 
While treatment the mass of solvent acetonitrile molecules was subtracted from the total molecular 
weight, so that it was taken as 1208.448 g/mol. Solvent losses is a characteristic feature of 
Mn$_4$(hmp)$_6$ structures. Fresh crystals lose solvent with time making the molar weight out 
of control, so that the error may reach 12\%. Therefore, in the present paper the data for dry (aged)
crystals left in the open dry air for several months were used. 

One may see in Fig.~\ref{chiT-exp} that the position of the low temperature maximum of 
$\chi T$ depends on the applied magnetic field, B: it shifts to higher temperatures with 
increase of B. Such a behavior is common for most of high spin molecules and molecular 
magnets.~\cite{Gatteschi2006} The origin of the maximum and its position in particular is 
not related with the microscopic characteristics of the exchange-coupled core. 
There are few possible explanations of the low temperature behavior of
the experimental $\chi T$ curve.

First of all, in paramagnetic systems the Curie law is only observed at temperatures 
$k_B T \gg g\mu_B S H$. For $T \lesssim T_p = g\mu_B S H$ thermal 
excitations are ineffective in providing dynamic Boltzmann equilibrium on higher Zeeman levels 
and the lowest level becomes ``overpopulated''. In theory, it means that one cannot expand the 
Brillouin function in the Taylor series at $x \ll 1$ (where $x=g \mu_B S H/k_B T$) and 
get the Curie law. 
In practice, the data points will not lay on the equilibrium $\chi(T)T$ curve. For B=0.2 T,
where the maximum in $\chi(T)T$ is first clearly observable, $T_p=2.4$~K and hence for the
temperatures of order of $T_p$ or lower the deviations from the Curie law has to be
observed. This is exactly what is seen in Fig.~\ref{chiT-exp}. It's worthwhile mentioning 
that the stronger magnetic fields shift this region of the ``inapplicability'' of the Curie
law to higher temperatures.~\cite{Jeon2010}

Second, there is no phase transition from correlated state (with spins S$_{Mn^3+}$ = 2
and S$_{Mn^2+}$ = 5/2) to high spin state S=9 for the individual Mn$_4$(hmp)$_6$ complex 
as well as for other heterospin SMMs. Experimentally, there 
is no Curie temperature that can be measured. There is a temperature domain, in which a crossover from 
correlated paramagnetic to a high spin superparamagnetic state occurs in an individual tiny single 
crystal. In this domain both our Heisenberg model and dynamic $S = 9$ approach are not applicable. However, 
there is a blocking temperature $T_b \sim 1 \div 2$ K, below which every complex in a tiny crystal becomes 
an anisotropic quantum magnet. In the measurement starting from that low temperature, $T_b$, the magnetic 
response becomes quantitatively irreversible and dependent on magnetic history due to magnetic 
anisotropy of individual complex.
In particular, this leads to a divergence of field cooled (FC) 
and zero field cooled (ZFC) $\chi T$ curves, as it happens in spin glasses. The FC type curves 
prevail in published data. A typical sample mass of order of 10 mg requires fields 1000 G and 
higher to get satisfactory paramagnetic response at 300 K. For more than 40 various Mn$_4$(hmp)$_6$ 
systems listed in Ref.~\onlinecite{Roubeau2008,Jerzykiewicz2010,Lecren2005} the experimental 
$\chi(T)T$ data were obtained at $B>0.1$ T, so that 
a sharp peak was present on every curve. In the experimental protocols $\chi(T)$ measurements 
are performed at heating from helium to ambient temperatures. They usually follow after magnetization 
field measurements, $M(B)$, where the excursions to high fields B=5 to 7 T take place. 
The dependence of that type inherits magnetic history of the sample and 
imports excessive (or deficient) magnetization from low to higher temperature region. 
This makes a broader temperature domain inappropriate for numerical analysis.

Third, the magnetic fields as low as $\sim$0.1 T are capable to produce a texture in thin 
polycrystalline magnets. For textured 
samples, the product  $\chi(T) T$ reaches maximum at higher temperatures. Ignoring the anisotropy 
the maximal $\chi(T) T$ values in high spin state are limited by 45 to 50 emu~K/mol for 
$g=2.0$ to 2.1 respectively. Note also that g-factor for $S=9$ state differs from its average 
value at high temperatures. As an example, a broad maximum of 42 emu K/mol at $T\approx$10 K 
was observed in polycrystals of  
[Mn$_2$Mn$_2$(teaH)$_2$(teaH$_2$)$_2$(O$_2$CPh)$_4$]$\cdot$0.7MeCN$\cdot$0.3EtOH nearly 
isostructural to the similar system not revealing it.~\cite{Ako2006} That might be an indication of a 
texture rather than drastic enhancement of exchange coupling. 
%No other dicubane Mn$_4$ systems 
%with the energy gap to first excited state (Ems=9-Ems=8) reaching 21 K were reported yet.

For analysis we select $\chi (T) T$ data points measured at B=0.01 T in order to avoid the 
the effect of the ``overpopulation'' of the lowest Zeeman level and reduce $T_p$ as described
above. Two sets of data ``at cooling'' and ``at heating'', presented in Fig.~\ref{chiT}, allow
to estimate the difference between these regimes. Down to 20 K both data coincide  
with the accuracy of 3.5\%. At 2.0 K the data 
diverge reaching 48.1 emu~K/mol ``at cooling'' and 53.9 emu~K/mol ``at heating''. The values 
14.3 emu~K/mol at 300 K are in agreement with other results reported on Mn$_4$(hmp)$_6$ structures 
and with the theoretical value 13.90 emu~K/mol for $g=$1.94. 
The value of the effective activation  barrier $\Delta_{eff}=|D_{eff}|S^2$ =17.9 K extracted 
from the ac data in Ref.~\onlinecite{Jiang2012} ($\Delta_{eff}$, $\tau_0$=4.89 10$^{-9}$s - parameters of Arrhenius law) fits in the range of published data usually found around 15-23 K (see Ref. 5-7 in 
Ako et al.~\cite{Ako2006}). This gives an estimate for the effective ZFS constant 
$D_{eff} =-0.2$ K in S=9 state. In the temperature domain 1.8$\div$3.5 K ($T>T_b$), the interplay of 
thermal and quantum relaxation processes of the magnetization 
occurs. The thermal barrier is therefore ``short-cut'' by the quantum tunneling of the magnetization 
and the obtained $D_{eff}$ value gives lowest estimate for the theoretical parameter. Upper estimate 
for $D_{S=9}$ may reach -0.4 K.

%%%%%%%%%%%%%%%%%%%%%%%%%%%%%%%%%%%%%%%%%%%%%%%%%%%%%%%%%%%%%%%%%%%%%%%%%%%
\section{Microscopic mechanism of the FM exchange coupling\label{Anal:sec}}
In order to understand why $J_1$ is much larger than $J_2$ and $J_{bb}$ 
one needs to find all the contributions to these exchange integrals. 
It is especially nontrivial to explain the difference between $J_1$ and $J_2$
because both constants describe
coupling between Mn$^{3+}$ and Mn$^{2+}$ ions in a similar geometry.

We will consider the superexchange coming from the virtual hopping of $d-$electrons via 
$p-$states of ligands, and use the 4th order of the perturbation theory.
The perturbation resulting to different energies of $\uparrow \uparrow$
and $\uparrow \downarrow$ states is given by the hopping integral $t$. 4th
order of the perturbation theory means that we are considering  only those
paths which consists of 4 hoppings and the initial and final states
are the same. The energy difference between initial and excited states 
(which appear while electrons hop) define the denominators in 
Eqs.~(\ref{t2gt2g:exchange}-\ref{t2geghf2:exchange}). The numerators
in Eqs.~(\ref{t2gt2g:exchange}-\ref{t2geghf2:exchange}) are given
by the corresponding hopping integrals $t_{pd}$ between the ligand $p$ and Mn $d$ orbitals
with different coefficients, which take into account the symmetry of the hoppings,
number of the hoppings of a given type etc. The detailed
description of this method can be found 
elsewhere.~\cite{Goodenough,Pavarini2012,Khomskii-book,ziese2001spin}

One needs to introduce the following parameters to find explicit expressions
for the exchange parameters: $U$ - is the on-site
Coulomb repulsion parameter, and $\Delta_{CT}$ - is the charge transfer
energy (energy of the excitation from the ligand $2p$ orbitals to the 
$3d$ shell of a transition metal ion). In general $\Delta_{CT}$ 
depends both on the valence state of the transition metal ion 
and type of the ligand~\cite{Bocquet1992,Ushakov2011}. For instance,
according to the result of T. Mizokawa the charge transfer
energy for Mn$^{2+}$ is $\Delta_{2+} \sim 7$ eV, which is much
larger than in the case of the Mn$^{3+}$ ions, for which  
$\Delta_{3+} \sim 4$ eV.~\cite{Ushakov2011}

\begin{figure}[t!]
 \centering
 \includegraphics[clip=false,width=0.3\textwidth]{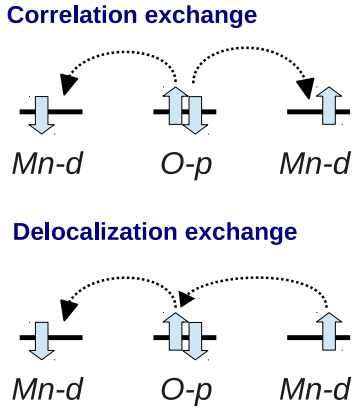}
\caption{\label{CorDel}(color online).
Two types of the contributions to the superexchange
according to the notations of Ref.~\onlinecite{Goodenough}.
}
\end{figure}

According to the terminology of Ref.~\onlinecite{Goodenough} there
are two important types of the exchange processes related with 
delocalization and correlation effects. They are sketched in Fig.~\ref{CorDel}.
The delocalization contribution to the exchange interaction involves 
the transfer of the ligand $p$ electron to one of the Mn ions, while 
$d$ electron from another Mn occupies the vacant place in the ligand 
$p$ shell (after that both electrons must return to
their initial places). The correlation effects are related with the transfer 
of two $p$ electrons to two $d$ sites on the right and on the left 
and then back.

For the simplicity in the analysis we will neglect the crystal-field splitting
between $t_{2g}$ and $e_g$ orbitals (which in general is not small,
$\sim$ 2 eV, in the case of the transition metal ions) and
the splitting of the Mn$^{3+}$ $e_g$ levels due to the
Jahn-Teller effect. These terms will effectively increase the denominators
in Eqs.~(\ref{t2geghf:exchange}-\ref{t2geghf2:exchange}).

\subsection{Delocalization contribution to $J_1$ and $J_2$\label{delexch}}
There are two special features of the studied Mn$_4$ molecular magnet which 
must be taken into account to find the expression for the delocalization
contribution to the $J_1$ and $J_2$ exchange constants.

First of all the electron transfer from Mn$^{3+}$  to Mn$^{2+}$ (back and forth)
is quite different from the transfer from Mn$^{2+}$ to Mn$^{3+}$ 
(back and forth). Neglecting the intra-atomic Hund's rule coupling 
($J_H$) in the first case the energies of the excited states 
(with respect to the energy of the unperturbed state) are $\Delta_{2+}$, $2U$ and 
again $\Delta_{2+}$. So that this electron transfer is highly unfavorable,
because it costs $2U$, where $U\sim$4.5-8 eV.~\cite{Streltsov2008,Medvedeva2002}
In contrast the second type of the electron transfer (from Mn$^{2+}$ to Mn$^{3+}$ back
and forth) cost neither $2U$ nor even $U$, one only needs to spend the energy $V$. 
$V$ is the energy of transfer (excitation) of an electron from Mn$^{2+}$ to Mn$^{3+}$, 
determined by the local coordination of these ions and by the intersite Coulomb interaction. 
Note that this energy $V$ is much smaller than the on-site Coulomb (Hubbard) repulsion $U$, therefore the 
process of the virtual transfer of an electron from Mn$^{2+}$ to Mn$^{3+}$, leading to 
superexchange between these ions, costs much less energy than that between the similar Mn 
ions, Mn$^{2+}$-Mn$^{2+}$ and Mn$^{3+}$-Mn$^{3+}$, and also less than the transfer in the 
pair Mn$^{2+}$-Mn$^{3+}$ in opposite direction, which would correspond to the ``reaction'' 
(Mn$^{2+}$,Mn$^{3+}$) $\to$ (Mn$^{1+}$,Mn$^{4+}$). Correspondingly, this process of virtual 
hopping from Mn$^{2+}$ to Mn$^{3+}$ would give largest exchange, which agrees with 
our numerical results.

\begin{figure}[t!]
 \centering
 \includegraphics[clip=false,width=0.3\textwidth]{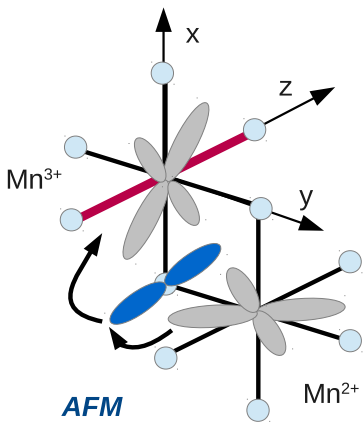}
\caption{\label{t2gt2g:fig}(color online).
The pair of the $t_{2g}$ orbitals participating in the superexchange 
interaction between Mn$^{3+}$ and Mn$^{2+}$ in the case of the $J_1$.
There will be the second pair for same ions, when the orbitals 
are interchanged and hopping occurs via the another common ligand. 
The ligands are shown as light blue circles, Mn $3d$ orbital
in grey, while ligand $2p$ orbitals in blue. Here and in Fig.~\ref{t2gegJ1:fig}-\ref{t2gegJ22:fig} 
arrows show the how the electrons move from one Mn to another.
}
\end{figure}

\begin{figure}[t!]
 \centering
 \includegraphics[clip=false,width=0.5\textwidth]{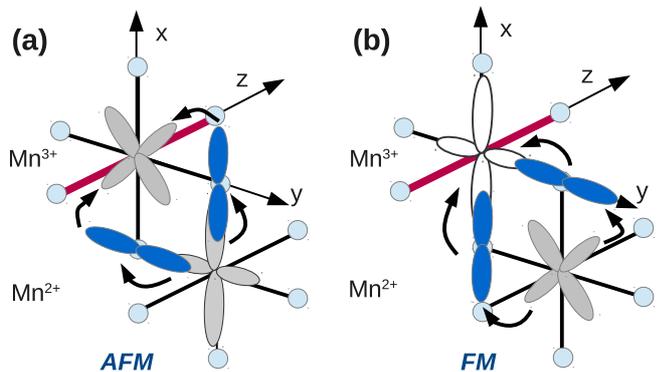}
\caption{\label{t2gegJ1:fig}(color online).
Two $t_{2g}/e_g$ contributions to the superexchange 
interaction between Mn$^{3+}$ and Mn$^{2+}$ in the case of the $J_1$.
The antiferromagnetic contribution is presented in the left panel (a),
while ferromagnetic in the right panel (b).
The ligands are shown as light blue circles, half-filled Mn $3d$ orbital
in grey, empty $x^2-y^2$ orbital in white, ligand $2p$ orbitals in blue. 
The long Mn$^{3+}$-O is shown in red.
}
\end{figure}

\begin{figure}[t!]
 \centering
 \includegraphics[clip=false,width=0.5\textwidth]{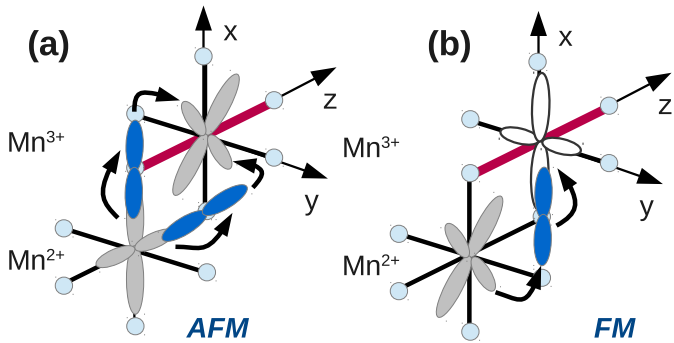}
\caption{\label{t2gegJ2:fig}(color online).
Two $t_{2g}/e_g$ contributions to the superexchange 
interaction between Mn$^{3+}$ and Mn$^{2+}$ in the case of the $J_2$.
The antiferromagnetic contribution is presented in the left panel (a),
while ferromagnetic in the right panel (b).
The ligands are shown as light blue circles, half-filled Mn $3d$ orbital
in grey, empty $x^2-y^2$ orbital in white, ligand $2p$ orbitals in blue. 
The long Mn$^{3+}$-O is shown in red.
}
\end{figure}

\begin{figure}[t]
 \centering
 \includegraphics[clip=false,width=0.3\textwidth]{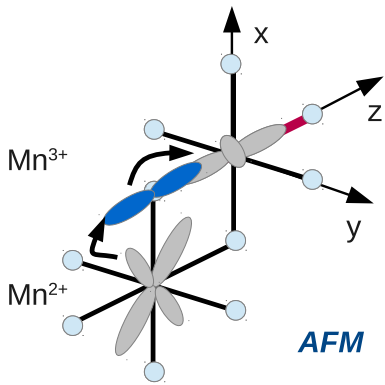}
\caption{\label{t2gegJ22:fig}(color online).
The antiferromagnetic $t_{2g}/e_g$ contributions to the superexchange 
interaction between Mn$^{3+}$ and Mn$^{2+}$ in the case of the $J_2$.
The ligands are shown as light blue circles, half-filled Mn $3d$ orbital
in grey,  ligand $2p$ orbitals in blue. 
The long Mn$^{3+}$-O is shown in red.
}
\end{figure}

The inter-site Coulomb interaction for Mn was estimated to be $\sim$0.5~eV
using constrained random-phase approximation.~\cite{Miyake2008} The
constrained LDA calculation for charge ordered Fe$_3$O$_4$ (where the 
number of the $d$ electrons is just slightly larger than in our situation)
gives $V=0.18$~eV.~\cite{Anisimov1996} One may expect
that $V \ll 2U$ in our case of Mn$_4$(hmp)$_6$ as well. In effect the second type of the 
electron transfer (i.e. from Mn$^{2+}$ to Mn$^{3+}$) will dominate and one may neglect the 
first type of the transfer (from Mn$^{3+}$ to Mn$^{2+}$). This is shown in 
Figs.~\ref{t2gt2g:fig}-\ref{t2gegJ22:fig} by arrows.

Moreover, the electron transfer from Mn$^{2+}$ to Mn$^{3+}$ back and forth costs 
also less charge transfer energy, since one needs to spend  $\Delta_{3+}$, and
as it was mentioned above $\Delta_{3+} < \Delta_{2+}$. Thus, we see
that the charge order strongly modifies the electron transfer
processes (and exchange processes as it will be shown below) and
should be explicitly taken into account.

Secondly, since the $3d$ shell in the case of the Mn$^{2+}$ is half-filled and
in Mn$^{3+}$ is close to half-filling, i.e. to the situation where the energy
gain due to the intra-atomic exchange coupling is maximal, one needs to properly
count the number of the Hund's rule constants $J_H$ for each electron transfer
process. As we will see below this will additionally stabilize ferromagnetic 
contributions to the exchange coupling between Mn$^{3+}$ and Mn$^{2+}$.

We start with the calculation of the contributions coming from exchange coupling 
between $t_{2g}$ orbitals on Mn$^{3+}$ and Mn$^{2+}$ ions. This term
is nearly the same for $J_1$ and $J_2$, since all considered orbitals are half-filled
on all sites, while the angle dependence of the hoppings can be neglected in the first approximation:
\begin{eqnarray}
\label{t2gt2g:exchange}
J^{t_{2g}/t_{2g}}_{1,2} = \frac {2t_{pd \pi}^4} {\Delta^2_{3+}(V+4J_H)}
\equiv J_0.
\end{eqnarray}
This contribution is antiferromagnetic and $J_0>0$. The factor 2 appears
because there are two pairs of the $t_{2g}$ orbitals, which take part in this 
superexchange process (one of the pairs for $J_1$ is shown in Fig.~\ref{t2gt2g:fig},
while the another one will act via the second common oxygen).

The $e_g/e_g$ contribution in the shared edge geometry is expected 
to be small, since the electrons are supposed to hop via almost 
orthogonal ligand $2p$ orbitals,~\cite{Streltsov2008}
and will not be considered here. In contrast the cross terms
from the $t_{2g}$ and $e_g$ orbitals are of the great importance.
They will be different for $J_1$ and $J_2$ because the single half-filled $e_g$ orbital
of Mn$^{3+}$ is directed differently in the pairs providing $J_1$ and 
$J_2$ exchange couplings.

There are two types of the $t_{2g}/e_g$ contributions. One is the hopping from 
the half-filled $e_g$ orbital of Mn$^{2+}$ to the half-filled $t_{2g}$ 
states of Mn$^{3+}$ ($t_{2g} \to e_g$) and back, see Fig.\ref{t2gegJ1:fig}(a). 
These terms are antiferromagnetic and the same for both $J_1$ and 
$J_2$. However there is also the ``opposite'' process, hopping of 
$e_g \to t_{2g}$, from Mn$^{2+}$ to Mn$^{3+}$ and back. Due to Jahn-Teller 
character of the Mn$^{3+}$ ion, with its particular orbital occupation (one 
$e_g$ electron of Mn$^{3+}$ occupies $3z^2-r^2$ -orbital, where the local $z-$axes 
is directed along the long Mn-O bonds, see Fig.~\ref{cryst.str} and red bonds in 
Figs.~\ref{t2gegJ1:fig},\ref{t2gegJ22:fig}), the contribution of this process 
would be different for $J_1$ and $J_2$. This is explained in details below.

We start with $J_1$. If the local $z$ axis is directed along the
longest Mn$^{3+}$-O bond, then the $x^2-y^2$ orbital of Mn$^{3+}$ 
must be empty, while the $3z^2-r^2$ orbital is half-filled. 
The expression for the exchange coupling between the half-filled $x^2-y^2$ 
orbital of Mn$^{2+}$ and the half-filled $xy$ orbital of Mn$^{3+}$
is very similar to Eq.~\eqref{t2gt2g:exchange} with the only difference
that here one of the $t_{pd}$ hoppings is of the $\sigma$ symmetry.
Since according to Ref.~\onlinecite{Slater1954} the hopping between
the $x^2-y^2$ and $p_x$ orbital is $(\sqrt{3}/2)t_{pd\sigma}$
in the given geometry and $t_{pd\sigma} \approx 2t_{pd\pi}$~\cite{Harrison1999}:
\begin{eqnarray}
\label{t2geghf:exchange}
J^{xy/x^2-y^2}_1 = \frac 32 \frac {t_{pd \pi}^2 t_{pd \sigma}^2} 
{\Delta^2_{3+}(V+4J_H)} = 3J_0
\end{eqnarray}
(since exchange occurs via two oxygens the prefactor equals
$2(\sqrt 3/2)^2=3/2$).
The orbitals providing 
this contribution are shown in Fig.~\ref{t2gegJ1:fig}(a). The exchange
coupling given by Eq.~\eqref{t2geghf:exchange}
are antiferromagnetic.

The ferromagnetic contribution comes from the interaction between
the half-filled $xy$ orbital of Mn$^{2+}$ and the empty
$x^2-y^2$ orbital of Mn$^{3+}$, which are shown in  Fig.~\ref{t2gegJ1:fig}(b).
Finding the difference between total energies of ferromagnetic
and antiferromagnetic solutions in the perturbation
theory one gets that
\begin{eqnarray}
\label{t2geghfe:exchange}
J^{x^2-y^2/xy}_1 = -\frac 32 \frac {t_{pd \pi}^2 t_{pd \sigma}^2 4J_H} 
{\Delta^2_{3+}V(V+4J_H)} =
-\frac {12 J_0 J_H}{V}.
\end{eqnarray}
This type of the ferromagnetic terms described by the second
Goodenough-Kanamori-Anderson rule~\cite{Goodenough,Khomskii-book} are usually quite
small because instead of the small $V$ there appears $U$, which is much
larger. In effect these contributions $\sim 1/U^2$ are considerably smaller 
than the conventional antiferromagnetic 
superexchange, which is $\sim 1/U$. This is one of the reasons why
the insulating transition metal oxides are mostly 
antiferromagnets.~\cite{Khomskii-97} Moreover typically the
ferromagnets are the systems with the small $U$ (YTiO$_3$,~\cite{Streltsov2005}
K$_{2}$Cr$_{8}$O$_{16}$,~\cite{Toriyama2011} Ba$_2$NaOsO$_6$,~\cite{Erickson2007} and etc.).
Since, as was mentioned above, in the present system $V$ is less than 
$U$ this term turns out to be quite efficient. 
The multiplier $4J_H$, which appears in the numerator in 
Eq.~\eqref{t2geghfe:exchange} due to the high spin state of Mn ions, additionally 
increases this contribution.

The $t_{2g}/e_g$ contributions to the $J_2$ can be obtained in a
similar manner. The antiferromagnetic coupling between the $x^2-y^2$
orbital of Mn$^{2+}$ and $t_{2g}$ orbital is exactly the same
as in the case $J_1$ [the difference is only in the notations: here
the $xz$ orbital is the active one, see Fig.~\ref{t2gegJ2:fig}(a)] and 
described by Eq.~\eqref{t2geghf:exchange}. 

The ferromagnetic contribution is also similar, but the coefficients
in expression for its value will be different. Due to different direction
of the long Jahn-Teller Mn$^{3+}$-O bond only one of the lobes of the 
empty $x^2-y^2$ orbital of Mn$^{3+}$ ion will be directed towards
Mn$^{2+}$, see Fig.~\ref{t2gegJ2:fig}(b). As a result the ferromagnetic term will be just
\begin{eqnarray}
\label{mainFMterm}
J^{x^2-y^2/xy}_2 = -\frac 34 \frac {t_{pd \pi}^2 t_{pd \sigma}^2 4J_H} 
{\Delta^2_{3+}V(V+4J_H)} =
- \frac{6 J_0 J_H}{V}.
\end{eqnarray}

In addiction to the reduction of the ferromagnetic contribution for this 
pair \eqref{mainFMterm} as compared to \eqref{t2geghfe:exchange}, there appears for this 
bond (due to different direction of the long Mn$^{3+}-O$ bond) also an antiferromagnetic 
contribution from the hopping between  the half-filled $xz$ orbital of Mn$^{2+}$
and the half-filled $3z^2-r^2$ orbital of Mn$^{3+}$, shown in 
Fig.~\ref{t2gegJ22:fig}:
\begin{eqnarray}
\label{t2geghf2:exchange}
J^{3z^2-r^2/xz}_2 = \frac {2t_{pd \pi}^2 t_{pd \sigma}^2} {\Delta^2_{3+}(V+4J_H)}=4J_0.
\end{eqnarray}
Since an opposite hopping from $3z^2-r^2$-Mn$^{2+}$ to 
$xz$-Mn$^{3+}$ is also possible, the prefactor 2 appears in the last equation.
Thus one may see that the ferromagnetic $t_{2g}/e_g$ component 
of $J_2$ turns out to be suppressed due to a specific orbital order
induced by the Jahn-Teller distortions of the Mn$^{3+}$O$_6$
octahedra, while the antiferromagnetic one is enhanced.

Combining above mentioned contributions one obtains, that
\begin{eqnarray}
\label{J1}
J_1 &=& (4 - \frac{12J_H}V)J_0, \\
J_2 &=& (5 - \frac{6J_H}V)J_0.
\label{J2}
\end{eqnarray}
Thus $J_1$ is ferromagnetic if $V<3J_H \sim 2.7$~eV. The
estimations of $V$ mentioned in Sec.~\ref{delexch} definitely satisfied
this condition, so that $J_1$, as well as $J_2$, are expected
to be ferromagnetic and $|J_1|>|J_2|$. However a care should be taken
with $V$ calculated in Ref.~\onlinecite{Miyake2008,Anisimov1996}, since in the 
present consideration $V$ is not only the inter-site Coulomb repulsion, but 
it also includes the effects of the different local environment of the Mn$^{3+}$ and
Mn$^{3+}$ ions. In order to calculate $V$ directly one may use the constrain
calculations as proposed in Ref.~\onlinecite{Anisimov1996}. This
lies beyond the scope of the present paper. Here we only provide
the upper limit for the value of $V$.

One may extract the unscreened value of $V$, recalculating it as
the center of the gravity difference for the $3d$ bands of Mn$^{3+}$ and 
Mn$^{2+}$ in the LDA approximation. Since there is a different number of the $d$ electrons
on these two ions there will be different contributions from the 
on-cite Coulomb repulsion $U$ to these orbital energies. This correction
can be written as $(U-J_H)(n_d-1/2)$~\cite{Anisimov1997} in the case Mn ions, where $n_d$ is
the number of the $d$ electrons per ion. Taking into account this correction, we 
obtained the unscreened value of $V \sim 2$~eV, so that even in this
situation $J_1$ must be ferromagnetic. However, according to the Koopmans'
theorem the orbital energies cannot be considered as excitation energies
(in which we are interested), but are subjected to the orbital relaxation
and electron correlation effects, which are the essence of the screening processes
and which can be quite efficient.~\cite{Solovyev2005}

We would like to note that presented above expressions for the exchange 
integrals can only be used for the qualitative understanding of the exchange processes 
in the Mn$_4$ molecular magnet. The ferromagnetic contributions from the
overlap between empty and half-filled orbitals would be reduced by three
factors. First of all one needs to take into account the crystal
field splitting ($\Delta_{CFS}$) between the $t_{2g}$ and $e_g$ shells in an appropriate way,
which will modify the denominators in 
Eqs.~\eqref{t2geghf:exchange}-\eqref{t2geghf2:exchange} (e.g. in 
Eq.~\eqref{t2geghf:exchange} one will need to substitute
$\Delta_{3+} \to \Delta_{3+} + \Delta_{CFS}$). Second, as is shown 
in Appendix~\ref{CorCon} there are antiferromagnetic terms related with the 
correlation contribution to the exchange coupling and these terms
are especially important in the case of $J_2$. Third, there will also be
an additional contribution coming from the antiferromagnetic interaction between 
the half-filled $t_{2g}$ orbitals and the ``belt'' of the $3z^2-r^2$ orbital 
(i.e. $r^2$ part) through $p$ orbital. Corresponding hopping is not small
and equals $t_{pd\sigma}/2$.~\cite{Slater1954} This will provide
additional contributions to Eqs.~\eqref{t2geghf:exchange},\eqref{t2geghfe:exchange}
and \eqref{t2geghf2:exchange} and modify Eqs.~\eqref{J1} and \eqref{J2},
but still leave the qualitative description of the exchange processes
correct.

Finally it is worth to mention that for the quantitative 
estimation of different contributions one needs to know exact values of the 
model parameters such as $J_H$  and $V$.

\subsection{$J_{bb}$ exchange}
The same delocalization and correlation effects will be
important for the exchange coupling between two Mn$^{3+}$,
i.e. for the $J_{bb}$ exchange constant. However, the sign and the value of 
the total exchange interaction may strongly depend on the details of the 
crystal structure: the Mn--O, Mn--Mn distances and the Mn-O-Mn bond angle.
The detailed analysis of the $J_{bb}$ does not seem to add much here,
since first of all it represents the usual superexchange consideration
for two Mn$^{3+}$ ions, which can be found elsewhere (e.g. in 
Ref.~\onlinecite{Streltsov2008}), and, second, the value of $J_{bb}$
is quite small. 

\section{Conclusions}
In the present paper we performed {\it ab initio} band structure calculations
for the Mn$_4$(hmp)$_6$ molecular magnet within the density
functional theory (DFT) using the GGA+U approximation.
The exchange parameters for the Heisenberg model were extracted from the
total energy calculations of several collinear spin configurations.
In contrast to a common belief, one of the exchange constants for two
pairs of the Mn$^{3+}$ and Mn$^{2+}$ ions (so called $J_{wb}$)
turns out to be the largest $J_{1}=-6.3$~K. Two other exchange
couplings are $J_2=-$0.5~K (another two pairs of Mn$^{3+}$ and Mn$^{2+}$
ions) and $J_{bb}=-0.3$~K (between the Mn$^{3+}$ ions).

The microscopic analysis based on the fourth order perturbation 
theory allowed to establish the mechanism of the strong exchange
coupling along the $J_1$ exchange path. Conventional superexchange
between two Mn ions in the edge sharing geometry is enhanced in 
Mn$_4$(hmp)$_6$ by the charge order. The charge disproportionation
leads to the situation, in which the lowest virtual excitations, contributing 
to the superexchange, will be not those across the Hubbard gap $\sim U$, but will 
be those between Mn$^{3+}$ and Mn$^{4+}$, which cost much smaller energy: the energy 
$V$ ($\ll U$) stabilizing the charge ordered state.
As a result the exchange coupling
between the empty $x^2-y^2$ orbital of Mn$^{3+}$ and the half-filled
$t_{2g}$ orbitals of Mn$^{2+}$, according to the second 
Goodenough-Kanamori-Anderson rule~\cite{Goodenough}, turns out to be 
quite effective and stabilizes ferromagnetic coupling along the $J_1$ exchange paths. 

In addition to charge order, there is an orbital order in Mn$_4$(hmp)$_6$,
which also has influence on the exchange interaction in this system.
The direction of the long Jahn-Teller Mn-O bond in the Mn$^{3+}$O$_6$ 
octahedra defines the orientation of the empty $x^2-y^2$ orbital.
This in turn regulates the absolute values of the exchange coupling between
different Mn$^{3+}$ and Mn$^{2+}$ pairs making one of them ($J_1$) 
larger than the other ($J_2$).

It is also important that the energy of the first excited state 
in the exchange process for the ferromagnetic state is reduced 
by a strong intra-atomic Hund's rule exchange coupling. This is
a feature of the Mn$^{3+}$ ion with $d^4$ electronic configurations
that has one empty $3d$ orbital and the energy difference between 
$(d\uparrow)^5$ and $(d\uparrow)^4 (d \downarrow )^1$ states is $4J_H$.
For any other configuration ($d^3$, $d^2$ etc.) this energy difference
between the excited states, according and against Hund's rule, will be 
smaller.

The exchange constants calculated in the GGA+U approximation
were used for the solution of the quantum Heisenberg model
for the given geometry. The magnetic susceptibility obtained by 
the exact diagonalization method reasonably agrees with the experimentally 
observed data. This additionally supports the results obtained by 
the DFT methods. The agreement between theoretical and experimental
data may be further improved by ``educated'' fitting, i.e. fitting, which
uses exchange constants obtained in the GGA+U approximation as a starting
point. The account of the single ion anisotropy on the Mn$^{3+}$ sites
also makes the agreement better. Although the exchange constants obtained by this fitting
are somewhat different from those calculated in the GGA+U
($J_{bb}^{fit,D}=-0.3$~K,  $J_{1}^{fit,D}=-3.6$~K, $J_{2}^{fit,D}=-0.4$~K, $D^{fit,D}=-0.2$~K). 
the general tendency is the same: $J_1$ is the largest exchange coupling.

One of the questions, which arises is whether it is possible to increase the
values of the ferromagnetic exchanges in the Mn$_4$ molecular magnets,
and what recipes one may provide on the basis of the microscopic model
developed in the present study. The substitution of all Mn ions
by Co, Ni, Cu ions would result in absence of the empty
$e_g$ orbitals, while changing Mn on Ti or V one reduces 
the energy gain due to the Hund's rule coupling. From this point of view the 
pair Cr$^{3+}$ ($t_{2g}^3$) and Cr$^{2+}$ (Jahn-Teller ion 
$t_{2g}^3 e_g^1$) could look more promising, if one could stabilize 
Cr$^{2+}$ (which is usually not easy). One may also expect some variations
of the exchange constants if the Fe$^{4+}$ and Fe$^{3+}$ instead
of the Mn$^{3+}$ and Mn$^{2+}$ ions will be used, since the intersite Coulomb 
repulsion, which contributes significantly to the energy of the
charge ordered state stabilization was reported to be quite small
for one of the Fe compounds.~\cite{Anisimov1996} The substitution of the
rare-earth elements instead of the transition metal ions may
lead to a significant gain in the Hund's rule energy, but 
it will simultaneously decrease the hopping integrals $t$.

Alternatively one may try to use not $3d$, but $4d$ or $5d-$elements. These typically have 
low-spin states with (most of) electrons in $t_{2g}$ shell and with $e_g$ states 
often empty and much smaller $U$ (than $3d$ counterparts). Then one could 
expect enhancement of ferromagnetic $t_{2g}$ $\to$ empty $e_g$ contribution. 
Unfortunately the crystal field splitting between $t_{2g}$ and $e_g$ levels 
for the $4d$ and $5d$ are usually larger and also the moment of respective ions are 
smaller than e.g. those of Mn$^{2+}$, Mn$^{3+}$.~\cite{Streltsov2012a,Zhou2012,Streltsov2013a}

Yet another option, which can be proposed, is 
the substitution of the
oxygen by the another ligand, e.g. sulfur. The charge transfer energy $\Delta_{CT}$
for the S$^{2+}$ ions is much smaller than for O$^{2+}$,~\cite{Bocquet1992} so that
if it would be chemically possible to substitute some of the
oxygens by sulphurs. This may lead to increase of the exchange 
coupling.

\section{Acknowledgments}
Authors thank D. Starichenko, who stimulated
the present study and participated in many interesting discussions
about the magnetic properties of the molecular magnets, A. Katanin and 
Prof. A. Mogilner for providing computational resources and support.

This work is supported by the Russian Foundation for Basic Research
(grants RFFI-13-02-00374, RFFI-13-02-00050, RFFI-12-02-31331,
RFFI 12-02-91371 the Ministry of education and science of Russia  
(grants 12.740.11.0026, MK-3443.2013.2 and FCP program), Samsung corporation via 
the GRO grant, by DFG via 1346 program and Research grant 1484/2-1, by
Cologne university via the German excellence initiative and by Russian Academy
of Sciences (programs 12M-23-2054, 12-P-2-1017, 12-CD-2). The calculations were 
performed on the ``Uran'' cluster of the IMM UB RAS.

\appendix
\section{Correlation contribution to $J_1$ and $J_2$~\label{CorCon}}
Very similar analysis can be performed for the correlation
contributions to the $J_1$ and $J_2$ exchange parameters. Below we briefly
discuss them and present some of the formulas.
 
Since as was mentioned in Sec.~\ref{delexch} the charge transfer energy 
from O $2p$ to Mn$^{2+}$ $3d$ shell ($\Delta_{2+}$) is much
larger than from O $2p$ to Mn$^{3+}$ $3d$ shell 
($\Delta_{3+}$) one may take into account only a part
of the correlation contributions related with the charge
transfer energy $\Delta_{3+}$, i.e. the processes, when
the first excitation occurs to the $3d$ shell of Mn$^{3+}$.
We will also neglect the terms describing the exchange coupling
via orthogonal $p$ orbitals, which are of order or 
$(t^4_{pd}/\Delta_{3+}^2)(J_H^p/(\Delta_{3+} + \Delta_{2+} + U_{pp})^2)$,
where $J_H^p$ is the Hund's rule coupling on the oxygen ion,
while $U_{pp}$ is the Coulomb repulsion between two holes
in the $p$ shell of oxygen.

The correlation contributions from the half-filled $t_{2g}$
orbitals will be antiferromagnetic and the same for $J_1$ and $J_2$:
\begin{equation}
J_{1,2}^{t_{2g}/t_{2g}} \sim
\frac{2t^{4}_{pd\pi}}{\Delta^2_{3+}}
\frac1{(\Delta_{3+} + \Delta_{2+} + U_{pp})}.
\end{equation}

In the case of $J_1$ the terms describing the correlation exchange between 
the empty $x^2-y^2$ orbital on the Mn$^{3+}$ ion and the half-filled 
$t_{2g}$ orbitals of the Mn$^{2+}$ and between half-filled 
$t_{2g}$ orbitals on the Mn$^{3+}$ and the half-filled 
$e_g$ orbitals of the Mn$^{2+}$ have different signs and 
nearly cancel each other out.

The direction of the Jahn-Teller distortion of the Mn$^{3+}$O$_6$ octahedra 
coincides with one of the Mn$^{3+}$--O bond forming the Mn$^{3+}$--Mn$^{2+}$ 
bond in the case of the $J_2$ exchange coupling. This upsets the delicate balance
between ferromagnetic and antiferromagnetic $t_{2g}/e_g$
terms and makes this contribution antiferromagnetic
\begin{equation}
J_{2}^{t_{2g}/e_{g}} \sim  \frac 34
\frac{t^{2}_{pd\pi}t^{2}_{pd\sigma}}{\Delta^2_{3+}}
\frac1{(\Delta_{3+} + \Delta_{2+} + U_{pp})}.
\end{equation}
This antiferromagnetic term additionally decreases the ferromagnetic
contribution to $J_2$ given in Eq.~\eqref{mainFMterm}.

\bibliography{../library}
\end{document}